\documentclass[twoside,a4paper,11pt]{sea9}
\usepackage{graphicx}
\usepackage{hyperref}
\begin{document}
\pagenumbering{arabic}
\pagestyle{myheadings}
\thispagestyle{empty}
{\flushleft\includegraphics[width=\textwidth,bb=58 650 590 680]{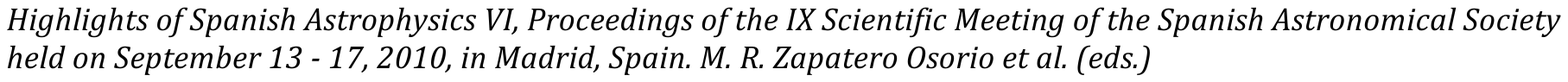}}
\vspace*{0.2cm}
\begin{flushleft}
{\bf {\LARGE
%
Exoplanet atmospheres: a brand-new and rapidly expanding research field
%
}\\
\vspace*{1cm}
%
Mercedes L\'opez-Morales$^{1,2}$
%
}\\
\vspace*{0.5cm}
%
$^{1}$
Institut de Ci\`encies de L'Espai (CSIC-IEEC), Campus UAB, Fac. Ci\`encies. Torre C5 parell 2, 08193 Bellaterra, Barcelona, Spain\\
$^{2}$
Carnegie Institution of Washington, Department of Terrestrial Magnetism, 5241 Broad Branch Road NW, Washington, DC, 20015, USA\\
%
\end{flushleft}
%
\markboth{
Exoplanet atmospheres
}{ 
%
Mercedes L\'opez-Morales
%
}
\thispagestyle{empty}
\vspace*{0.4cm}
\begin{minipage}[l]{0.09\textwidth}
\ 
\end{minipage}
\begin{minipage}[r]{0.9\textwidth}
\vspace{1cm}
\section*{Abstract}{\small
%
The field of exoplanets is quickly expanding from just the detection
of new planets and the measurement of their most basic parameters,
such as mass, radius and orbital configuration, to the first
measurements of their atmospheric characteristics, such as
temperature, chemical composition, albedo, dynamics and
structure. Here I will overview some the main findings on exoplanet
atmospheres until September 2010, first from space and just in the
past two years also from the ground.

%
\normalsize}
\end{minipage}
%
%
%
\section{Introduction \label{intro}}

The introduction of any general talk about exoplanets always includes
an account of the number of planets found. The number of planets
discovered as of mid-September 2010 was
490\footnote{http://exoplanet.eu}. All exoplanets detected thus far,
and recognized as such by the IAU, have been discovered within the
past 18 years via a series of observational techniques, which have
been quickly improving over time. That improvement is reflected in the
histogram in Figure 1, where one can not only see how the number of
exoplanets discovered per year has been steadily increasing,
but also the number of techniques with which exoplanets are being
discovered.

\begin{figure}
\center
\includegraphics[width=10cm,angle=0,clip=true]{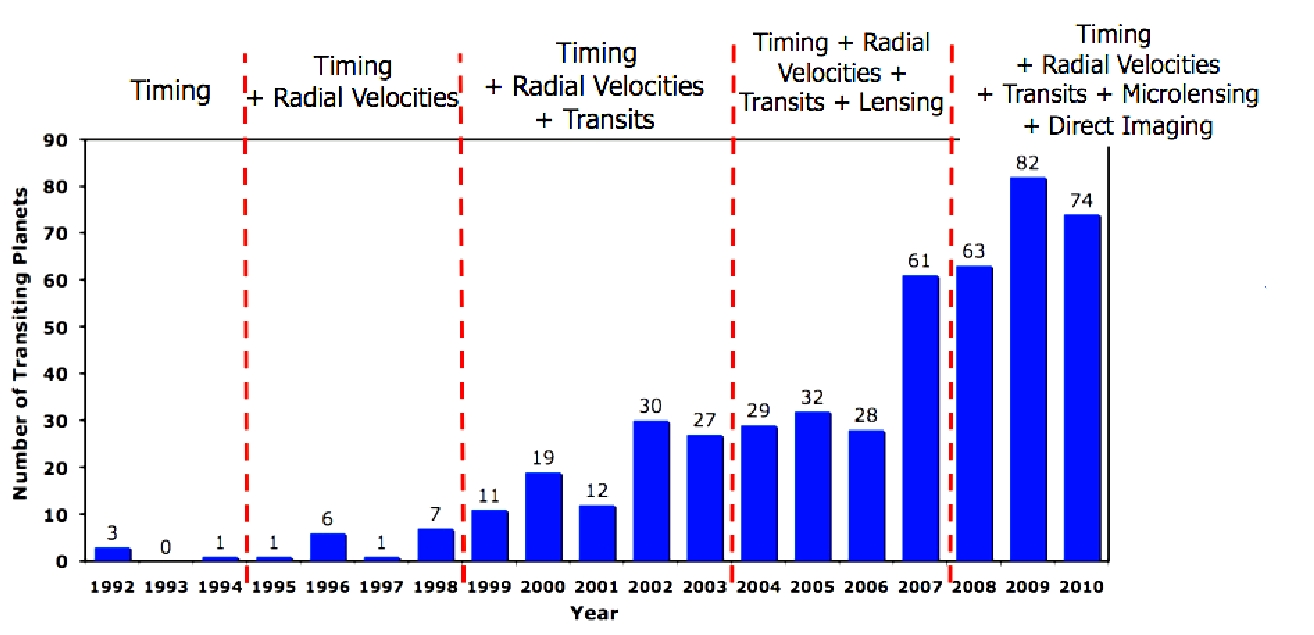}
\caption{\label{fig1} Histogram representing the number of exoplanets
  discovered per year since 1992 (last updated on
  September 2010; http://exoplanet.edu). On the top of the diagram are
  highlighted in chronological order the number of successful planet
  detection techniques. Each vertical dashed line indicates the year
  of the first discovery by each new technique, e.g. the first planet
  via the radial velocity technique was detected in 1995, and so on.}
\end{figure}

The first four exoplanets acknowledged as such were discovered between
1992 and 1994 via periodic timing variations in the very precise
signal of pulsars \cite{Wolszczan1992,Wolszczan1994}. Then, in 1995,
the first planet about a star like the Sun was discovered
\cite{Mayor1995}. This planet, similar in mass to our Jupiter and
orbiting its host star with a period of about four days, was
discovered via the radial velocity technique. This technique continues
to be the most successful, scoring close to 80$\%$ of the exoplanets
discovered to date. Three other techniques have accomplished the detection of
planets since then. In 1999 the discovery of the first planet via the
transits technique was announced
\cite{Charbonneau2000,Henry2000}. Then, in 2004, the micro-lensing
technique produced its first detection \cite{Bond2004}. Finally, just
two years ago, the first believed to be {\it bona-fide} imaged planets
were detected via direct imaging \cite{Kalas2008,Marois2008}.

With close to 500 planets discovered so far, just finding another
planet is no longer news, and we are witnessing a split of the field
of exoplanets into two main directions: on one hand work continues on
trying to detect new planets, but now focusing on finding the first
Earth analogs. On the other hand, many efforts are being put into
trying to unveil the physical properties of those planets, specially
the physical characteristics of their atmospheres. This is how a new
field, which some call {\it Exoplanetology}, is now being born.

\section{Exoplanetology}

As in other fields with a name ending in {\it -ology}, e.g. Biology,
Geology, and so on, Exoplanetology treats exoplanets like a group of
study {\it subjects} whose characteristics we want to understand in
detail. The information about exoplanets that can be obtained using
current techniques is definitely more limited than the information we
can derive from the technically more accessible planets in our own
Solar System. However, there are several properties of known
exoplanets that can be readily obtained.

From the timing, radial velocity and micro-lensing techniques we can
derive the minimum mass of the planets, $Msini$, where $i$ is the
inclination angle of the planet's orbit with respect to us as
observers. We can also measure the orbital period of the planet $P$ and 
its orbital separation from the host star $a$.

Transits and direct imaging provide a richer set of information. From
transiting systems we can measure the orbital inclination of the
system (and therefore the absolute mass of the planet whenever radial
velocities are available), the radius of the planet, $R_p$,
(and from that mass and radius we can derive the density of the planet
and determine if it is a rocky planet or a gas ball), the eccentricity
of the orbit, $e$, and we can also begin measuring some of the
planet's atmospheric characteristics. Via direct imaging we can
measure accurate orbital parameters of the system, as well as the
planetary atmosphere.

The atmospheric parameters that can be currently measured either through
transits or direct imaging are:

\begin{itemize}
\item the atmospheric temperature of the planet, $T_p$, as a function
  of wavelength, or equivalently, as a function of atmospheric height.
\item its bolometric albedo, $A_B$, which is the fraction of incident
  irradiation from the star on the planet's atmosphere that gets
  reflected back to space. This parameter gives us an idea of the
  presence or absence of clouds.
\item the energy re-distribution factor, $f$ or $P_n$\footnote{See
  Appendix in \cite{Spiegel2010} for a more detailed definition of
  these two parameters.}, which gives information about what fraction
  of the incident stellar energy on the irradiated day-side of the
  planet is transported to the non-irradiated night-side (see also
  Sections 4 and 5). From the energy re-distribution factor we can
  infer the presence or absence of winds in the atmosphere of the
  planet. To get a sense of the scale of these parameters, $f = 2/3$
  (or $P_n$ = 0) indicates completely inefficient transportation of
  energy between the irradiated and non-irradiated sides of the
  planets, while $f = 1/4$ (or $P_n$ = 0.5) indicates very efficient
  energy transportation.

\item and, finally, we can also start obtaining information about the
  chemical composition of the planet's atmosphere.
\end{itemize}

\section{Transiting planets}

As technology improves, direct imaging will eventually become the
technique of choice to characterize exoplanet atmospheres. In fact,
the first results using this technique have already been reported
\cite{Janson2010}, where from a spectrum between 3.88 and 4.10 $\mu$m
of the directly imaged planet HR8799c \cite{Marois2008}, the study
concludes that the continuum atmospheric emission from this planet
differs significantly from existing models. However, direct imaging is
still at its very early stages and, at present, transiting planets
provide the richest source of information about exoplanet atmospheres.
What makes a transiting planet so especial is that its orbital
inclination, as seen from Earth, is such that the planet crosses in
front and behind the star once every orbit, as illustrated in Figure
2a. The passage of the planet in front of the star is called a transit
(or {\it primary~eclipse}), while the passage of the planet behind the
star is usually called a {\it secondary~eclipse}.

\begin{figure}
\center
\includegraphics[width=10.5cm,angle=0,clip=true]{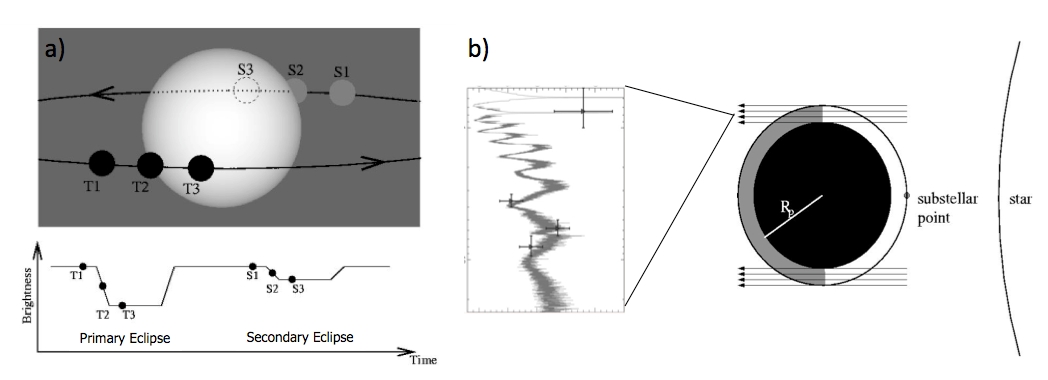}
\caption{\label{fig2} (a) Trasiting planet system, where once every
  orbit the planet crosses in front of the star (primary eclipse) and
  behind the star (secondary eclipse); (b) when a planet crosses in
  front of the star, part of the stellar light crosses through the
  atmosphere of the planet and is absorbed by the chemicals in that
  atmosphere. That extra absortion of stellar light produces what is
  called the {\it transmission~spectrum} of an exoplanet
  \cite{Tinetti2007}. }
\end{figure}

Both primary and secondary eclipses give a wealth of information about
the atmospheric characteristics of those planets, provided that we can
detect the signals, which are typically of the order of 1$\%$ of the
total light of the system for primary eclipses, and a small fraction
of a percent (0.1--0.2$\%$), for the secondaries.

In the case of primary eclipses, we can measure the chemical
composition of a planet's atmosphere, and the distribution of those
chemicals with atmospheric height, from what is called its {\it
  transmission~spectrum} (see Figure 2b).  When the planet passes in
front of the star, light from the star crosses the thin layers of its
atmosphere. The chemicals present in that atmosphere absorbe part of
the stellar light, and their signatures, i.e. their transmission
spectrum, can be identified by subtracting spectra of the system
collected during transit from spectra collected outside of transit.

Secondary eclipse observations provide an even more powerful tool. To
explain the concepts, although more detailed physical models are in
fact used in the interpretation of secondary eclipse data, as will be
shown later, we can assume that the planetary atmopshere behaves like
a blackbody. From the secondary eclipse we can measure directly the
flux coming from the atmosphere of the planet\footnote{in fact what we
  measure is the ratio of the planet-to-star fluxes, $F_p/F_{st}$, and
  then derive $F_{st}$ from models or real observations when
  available.}. Inserting that flux into Planck's equation
\begin{equation}
 F_P = \frac{2h\nu^3}{c^2} \frac{\pi R_{p}^2}{e^{\frac{h\nu}{kT_p}} -1} \frac{1}{D^2}~~; ~~\nu = c/\lambda,
\end{equation}

\noindent where we can estimate the blackbody temperature, $T_p$, on
the side of the planet that faces the star once we know the radius of
the planet, $R_p$, the distance of the planet-star system from Earth,
$D$, and the wavelength of the observations, $\lambda$. Then,
substituting that temperature into the energy balance equation

\begin{equation}
 T_P = T_{*} {\frac{R_{*}}{a}}^{\frac{1}{2}} [f(1-A_B)]^{\frac{1}{4}},
\end{equation}

\noindent where $R_{*}$, $T_{*}$, and $a$ are known, we can estimate
the bolometric albedo and the energy re-distribution factor of the
planet (see \cite{Lopez2007} for more details). Note that to do these
estimations we assume that $A_B \in [0,1]$ and $f \in [1/4,2/3]$.

From secondary eclipses it is also possible to obtain the {\it
  emission~spectrum} of the planet, either by measuring the spectrum
directly, or by building up a composite spectrum using photometric
observations from individual filters. However, spectroscopy is
preferred to minimize the potential effects of variability in the
planet's atmosphere (see end of section 4).

Until here the introduction of the concepts and terminology that will
be used in the remaining of the document. Let's now move to the main
observational results so far in the study of exoplanet
atmospheres. Notice that this is not a complete compilation of all
results, but a brief summary of the ones I consider have been most
relevant and illustrative. Notice also that this summary is based only
on results published until September 2010.

\section{Results from space}

From space we have already detected the atmosphere of about ten
planets, the majority of them are Jupiter-mass targets orbiting their
host star at very close distances, so-called {\it hot~Jupiters}. They
have the peculiarity of being tidally locked to their star, which
means that, like in our own Earth-Moon system, the same side of the
planet always faces the star. Therefore those planets have permanent
day and night sides. The most relevant detections from space have been:

\begin{itemize}
\item Via primary eclipses:
\begin{itemize}
\item The detection of Sodium (NaI) absorption in the atmosphere of
  the hot Jupiter HD 209458b, which was first reported in 2002
  \cite{Charbonneau2002}. These authors detected the sodium with the
  Hubble Space Telescope (HST), by observing an extra dimming during
  primary eclipse at the wavelengths corresponding to the NaI-doublet,
  around 0.59 $\mu$m. This was the first time the atmosphere of an
  exoplanet was detected.

\item The second detection of an exoplanet atmosphere was reported
  five years later \cite{Tinetti2007}. That study reported $H_2O$
  vapor in the atmosphere of another hot Jupiter, called HD 189733b,
  by measuring the transmission spectrum of the planet with the
  Spitzer Space Telescope (SST) at 3.6, 5.8 and 8.0 $\mu$m infrared
  wavelengths, and in the optical at $\sim$ 0.76 $\mu$m, with HST.
  The following year a new study \cite{Swain2008} claimed to not only
  see $H_2O$ features in the transmission spectrum of HD189733b, but
  also methane ($CH_4$). They arrived to this conclusions from HST
  spectra between 1.5 and 2.5 $\mu$m. Notice however, that there is
  being some controversy around these detection claims, as described
  at the end of this section.
\end{itemize}

\item Via secondary eclipses:
\begin{itemize}

\item atmospheric emission from the hot Jupiters HD 209458b
  \cite{Deming2005} and TrES-1b \cite{Charbonneau2005} with the
  SST. The detected emission of HD 209458b at 24 $\mu$m suggested a
  brightness temperature for the planet of about 1130 K. In the case
  of TrES-1b, the observed eclipse depths at 3.6 $\mu$m and 4.5 $\mu$m
  indicated a blackbody temperature of about 1060 K. These two were
  the first detections of atmospheric emission from exoplanets.

\item Also, using the SST, variations in brightness as the day and
  night sides rotate into view have been measured for two
  planets. Observationally, this effect is similar to the changes in
  brightness with phase seen in Venus, but while with Venus we can
  observe the planet directly, in the case of extrasolar planets, the
  light from the star is always on the way. One of the planets
  observed is {\it Ups~And~b}, which shows brightness changes between
  the day and the night side of the planet with an amplitud of about
  0.3$\%$ \cite{Harrington2006}\footnote{Notice that Ups And b,
    although included in this section, in fact does not eclipse, but
    it was still possible to measure the variation between its day and
    night side emission by observing it at five different epochs, when
    the planet was expected to go through different {\it illumination}
    phases and therefore contribute by different amounts to the total
    light of the system.}. For the other observed planet, HD189733b
  \cite{Knutson2007}, the change in brightness between the day and
  night sides is six times smoother, with an amplitude of only about
  0.05$\%$. Translating these differences in brightness into
  temperatures (assuming the planets emit as blackbodies), the
  atmosphere of Ups And b has a day-side temperature of about 1600K,
  while its temperature is $\sim$ 1000K cooler on the non-irradiated
  night-side. HD189733b, on the other hand, presents a more uniform
  temperature distribution, with a day-side at about 1200K, and a
  night-side only $\sim$ 240K cooler, or what is the same, HD189733b
  seems to very efficiently re-distribute the stellar incident energy
  around its atmosphere, while the same process seems very inneficient
  in Ups And b.

  This detection of the two outmost possible cases of stellar
  irradiated energy re-distribution efficiency at a so early stage in
  this field, has started a debate about whether there are two classes
  of irradiated hot Jupiters: one class where there is very little
  transport of irradiated energy to the night-side, and therefore the
  planets show very high brightness and temperature contrasts between
  their day and night sides, and another class where the energy gets
  quickly advected around the planets. A possible explanation for these
  effects is the presence, or absence, of {\it
    thermal~inversion~layers} in the upper atmosphere of those planets,
  similarly to the Ozone layer on the Earth's atmosphere, which
  absorbe and re-emit the incident stellar irradiation before it has
  time to penetrate into the planet's atmosphere and get
  advected. Ozone, however, cannot be responsible for those thermal inversion
  layers, because this molecule is unstable at the high temperatures
  observed in the upper atmospheres of hot Jupiters. The chemicals
  responsible for those thermal inversion layers remain unknown,
  although molecules such as TiO, VO or sulfur compounds have been
  suggested \cite{Fortney2008,Zahnle2009}.

\item The detection at optical wavelengths of the secondary eclipses
  of the hot Jupiters CoRoT-1b, CoRoT-2b and HAT-P-7b using the CoRoT
  and Kepler satellite missions
  \cite{Snellen2009,Snellen2010,Borucki2009}. As shown in Figure 3,
  the depths of the eclipses are of the order of $0.01\%$, and for two
  of the planets (CoRoT-1b and HAT-P-7b) one can nicely see curvatures
  in the light curve indicative of brightness variations with phase,
  which suggest that those planets have bright day sides versus
  significantly darker night sides and therefore might present thermal
  inversion layers in their atmospheres. The light curve of CoRoT-2b,
  on the other hand, appears flatter which can be interpret as a
  planet with more uninform temperature and no thermal inversion
  layer. An additional conclusion that can be derived from the very
  shallow optical depths of the eclipses is that all three planets
  have albedos close to $A_B = 0$.

\begin{figure}
\center
\includegraphics[width=10.5cm,angle=0,clip=true]{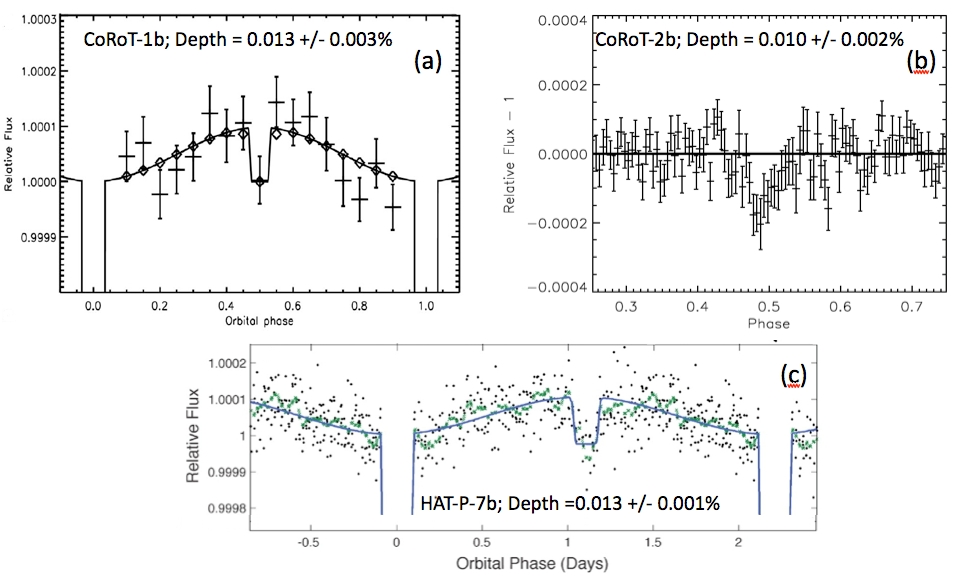}
\caption{\label{fig3} Secondary eclipses in the optical of the hot
  Jupiters (a) CoRoT-1b \cite{Snellen2009}, (b) CoRoT-2b
  \cite{Snellen2010}, both detected using the CoRoT Mission data, and
  (c) HAT-P-7b \cite{Borucki2009}, detected using Kepler Mission
  data.}
\end{figure}

\end{itemize}

\end{itemize}

The last set of results from space I present in this section is a
combination of primary and secondary eclipse measurements of
HD189733b, in the optical and near-infrared, to highlight the
potential discovery of variability in the atmosphere of that
planet. Shortly after \cite{Tinetti2007} announced the detection of
$H_2O$ vapor in this planet via transmission spectra, another paper
\cite{Grillmair2007} claimed no sign of water in the emission spectrum
of the planet during secondary eclipse. Another four papers had been
published by the end of 2009, alternating water/no-water claims in the
atmosphere of the planet
\cite{Swain2008,Pont2008,Grillmair2008,Sing2009b}. In a follow-up
result the same group who had claimed no water detection reported new
observations of the emission spectrum of the planet where they claimed
strong w≈ater absorption features and attribute the previous
non-detection to weather patterns in the atmosphere of the planet,
which would cause its emission spectrum to vary over time
\cite{Grillmair2008}. The most recent result reports the absence of
water absorption bands \cite{Sing2009b}. Instead, their data appear
most consistent with a featureless spectrum like the one previously
measured in the optical \cite{Pont2008}, and was attributed to some
kind of haze in the upper atmospheric layers of the planet. This is an
example of the kind of controversial findings that are starting to
emerge in this novel field. Either the atmosphere of the planet is
really varying, or perhaps what we are observing is stellar
variability or systematics between the observational strategies
and analyses performed by each group.

\section{Results from the ground}

The first successful detections using ground-based telescopes occured
in 2008 and were done via transmission spectra during primary eclipse,
again around the NaI doublet lines.

In the first detection, published with data from the 9.2-m
Hobby-Eberly Telescope (HET)\cite{Redfield2008}, they detected NaI
absorption in the $\sim0.59~ \mu m$ doublet in the transmission spectrum
of HD189733b. The signal was very small, only about 0.3$\%$ fainter
than the spectral continuum, but they managed to produce a larger than
3$\sigma$ detection. HD189733b is now the second planet with NaI
detected in its atmosphere.

In the second detection \cite{Snellen2008}, they revealed the NaI
doublet in the transmission spectrum of HD209458b with a confidence
level better than 5$\sigma$, using data from the 8.2-m Subaru
Telescope. Although this was not a new finding, but a confirmation of
the 2002 detection with HST \cite{Charbonneau2002}, it represented a
significant improvement on the detection of exoplanet atmospheric
transmission features using ground-based telescopes.

The third detection not only found carbon monoxide (CO) in the
atmosphere of HD209458b, but also found that the CO absorption to be
Doppler shifted by about 2 $kms^{-1}$(2$\sigma$), which suggests the
presence of strong longitudinal winds between the irradiated and
non-irradiated hemispheres of the planet \cite{Snellen2010b}.

The fourth, and last detection via transmission spectra was the
identification of potassium absorption around 0.76 $\mu$m in the
atmosphere of two different planets, XO-2b \cite{Sing2010} and
HD80606b \cite{Colon2010}. Absorption by Potassium, together with
Sodium, had been predicted ten years ago to be the most
prominent features in the transmission spectrum of exoplanets
\cite{Seager2000}.

Those four are, thus far, all the available detections of exoplanetary
atmospheres from the ground via primary eclipses. Detections of
planetary atmospheric emission via secondary eclipses have, on the
other hand, rocketed since early 2009.

The first two ground-based detections of thermal emission from
exoplanets were simultaneously announced in January 2009.  The first
of the two results reported the detection of the secondary eclipse of
the hot Jupiter OGLE-TR-56b, observed with the 6.5-m Magellan
Telescopes and the 8.2-m VLTs, in z'-band ($\sim$0.9 $\mu$m),
with a depth of $0.036 \pm 0.009 \%$ \cite{Sing2009a}. The other
detection, using the 4.2-m WHT, was also of a hot Jupiter, TrES-3b,
but in the near-IR K-band ($\sim$2.2 $\mu$m), where they reported a
depth of $\sim0.24 \pm 0.04 \%$ \cite{deMooij2009}.

Both detections are consistent with the irradiated atmospheres of the
planets being hotter than 2000K, and with the planets having a very
hot day-side versus a much colder non-irradiated night-side. This
means that, for both planets, the re-circulation of energy from the
day to the night side is very inefficient. The results are therefore
consistent with both planets having thermal inversion layers in their
upper atmosphere, similarly to the results found by SST for several
other planets. However, these observations in the optical and near-IR
prove different atmospheric depths (or pressure levels) than the SST
observations.

\begin{figure}
\center
\includegraphics[width=9.4cm,angle=0,clip=true]{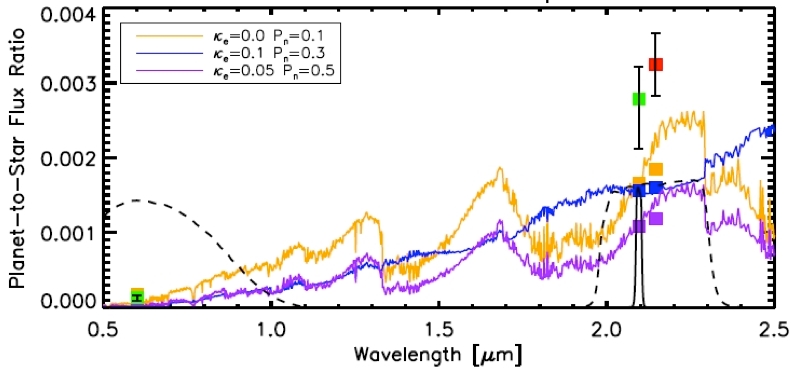}
\caption{\label{fig4} CoRoT-1b secondary optical and near-IR
  detections (green and red points), compared to models with
  different energy re-distribution parameters, $P_n$, and
  opacities, $\kappa_e$, which simulate the effect of thermal
  emission layers. None of the models, which assume the inversion
  layers at a height pressure of 1 mbar, can reproduce the
  two near-IR observations \cite{Rogers2009}.}
\end{figure}

Other secondary eclipse detections since then have been:
\begin{itemize}

\item The third ground-based detection of emission from an exoplanet's
  atmosphere was announced in June 2009 \cite{Gillon2009}. The planet
  in this case was CoRoT-1b, and they detected its secondary with the
  VLT, using a narrow-band filter centered at 2.09 $\mu$m. The depth
  of the eclipse that they measured was $\sim 0.278 \pm 0.054 \%$.
  Shortly after, the detection of that secondary was confirmed in
  K-band ($\sim$2.2 $\mu$m) with observations from the 3.5-m telescope
  at Apache Point Observatory (APO) \cite{Rogers2009}. This new study
  not only detected a secondary $\sim0.336 \pm 0.042
  \%$ deep, but also combined their result with the previous narrow-band
  filter detection \cite{Gillon2009} and the detection in the optical
  described in Section 4 \cite{Snellen2009} to compare, for
  the first time, observations of the atmospheric emission of an
  exoplanet in the optical and near-IR to current theoretical
  models. The result of that comparison, shown here in Figure
  4, is that, while the optical flux of the planet
  fits the models fairly well, the near-IR flux is at
  least twice larger than predicted by models. Such high near-IR
  flux also exceeds what would be expected for a planet with $A_B = 0$
  and $f = 2/3$. Therefore, not only the atmosphere of CoRoT-1b seems
  to have thermal inversion layers, but also could have some emission
  mechanism which makes the irradiated side of the planet appear too
  bright in the near-IR.

\item Emission detections from several other planets
  have been announced in 2010:

\begin{itemize}
\item In early 2010 the detection of thermal emission in z'-band
  ($\sim 0.9~\mu m$) from the hot Jupiter WASP-12b was announced,
  using data from the 3.5-m in APO \cite{Lopez2010}. They detected an
  eclipse depth of $\sim0.082 \pm 0.015 \%$ and also claimed a slight
  eccentricity of the planet's orbit, in agreement with the
  eccentricity claimed in the dicovery paper
  \cite{Hebb2009}. Subsequent studies
  \cite{Campo2010,Husnoo2010,Croll2010a} have disputed that
  eccentricity result. The last of those works, \cite{Croll2010a},
  also detects the thermal emission of WASP-12b in the near-IR J, H
  and $K_s$-band, and concludes that the higher pressure, deeper
  atmospheric layers of the planets probed by the J and H-band
  observations seem to be cooler, and more temperature homogenized
  than the higher atmospheric layers probed by $K_s$-band, where the
  observed emission from the planet is consistent with inefficient
  day-to-night redistribution of the irradiated heat, and low albedo.

\item Shortly after, two independent detections of thermal emission
  from WASP-19b, one in H-band ($\sim 1.60~ \mu m$) and the other one
  using a narrowband NB2090 ($\sim 2.09~ \mu m$) filter where
  announced \cite{Anderson2010,Gibson2010}. Both results were obtained
  with the VLT. The depths of both eclipses, $\sim0.366 \pm 0.072 \%$
  in NB2090 and $\sim0.259 \pm 0.045 \%$ in H-band, are consistent
  with a very low albedo, and very inefficient atmospheric energy
  re-distribution of the irradiated stellar energy. In addition, the
  emission of the planet in both filters give an atmospheric
  temperature for the planet of about 2600K, which indicate that
  WASP-19b, likewise CoRoT-1b as explained before, is brighter in the
  near-IR bands than predicted by any existing model.

\item Results for another two planets, TrES-2b and TrES-3b, were
  reported in the Spring of 2010 \cite{Croll2010b,Croll2010c}. In
  the case of TrES-2b they detected an eclipse depth of $\sim0.062 \pm
  0.012 \%$ in $K_s$-band, which gives a day-side blackbody
  temperature for this planet of about $1600 \pm 100$K. This
  $K_s$-band emission, when combined with the SST secondary eclipse
  detections \cite{ODonovan2010}, suggests that the atmosphere of
  TrES-2b exhibits relative efficient day-to-night side
  re-distribution of heat. This is the first planet with a secondary
  eclipse observed from the ground for which an atmopheric thermal
  inversion layer might not be required to explain the observed flux.
  In the case of TrES-3b, these authors detected a $K_s$-band eclipse
  depth of $\sim0.133 \pm 0.017 \%$, which is about half the depth
  previously reported \cite{deMooij2009}. They also tried to detect the
  thermal emission of the planet in H-band and could only come up with
  a 3$\sigma$ upper limit of 0.051$\%$. This new $K_s$-band detection
  requires very efficient energy re-distribution in the atmosphere of
  TrES-3b. In addition, the authors speculate with the possible
  presence of a highly absorbent molecule, such as methane, to explain
  the non-detection of emission in H-band.

\end{itemize}

\end{itemize}

All the secondary eclipse detections described above have been done
photometrically, using either standard broadband or narrowband
photometric filters. However, the first result at the next level,
i.e. a spectro-photometric detection of emission from an exoplanet,
has been already announced using the median-size 3.0-m InfraRed
Telescope Facility (IRTF) \cite{Swain2010}. They detect emission from
the hot Jupiter HD189733b in two different spectroscopic windows,
between 2.0--2.4 $\mu$m and 3.0--4.1 $\mu$m, with an average spectral
resolution of R = 470. By binning these low resolution planetary
spectra into the equivalent of nine narrowband filters, they detect
eclipse depths from just a fraction of a percent, to a $\sim 1.0 \%$
depth in the bin centered around 3.3 $\mu$m. These results are
illustrated in Figure 5, where the authors compare the new
ground-based measurements to previous detections of emission from this
planet with the HST \cite{Swain2008} and the SST
\cite{Charbonneau2008}, and with {\it
  Local~Thermodynamical~Equilibrium} (LTE) models which reproduce well
those data. They find that the ground-based observations between 2.0
and 2.4 $\mu$m are both consistent with the HST observations and LTE
models. However, the spike observed around 3.2--3.3 $\mu$m cannot be
modeled by thermal emission alone. The suggested explanation is the
presence of non-LTE emission processes in the atmosphere of HD189733b,
which would cause methane fluorescence around those
wavelengths. Although this would be the first time fluorescent methane
emission is detected in an exoplanet, this effect has been seen in our
Solar System in the atmosphere of Jupiter, Saturn and
Titan\footnote{After this proceedings was submitted, a new result by
  \cite{Mandell2010} questions the emission features reported by
  \cite{Swain2010}.}.

\begin{figure}
\center
\includegraphics[width=8.0cm,angle=0,clip=true]{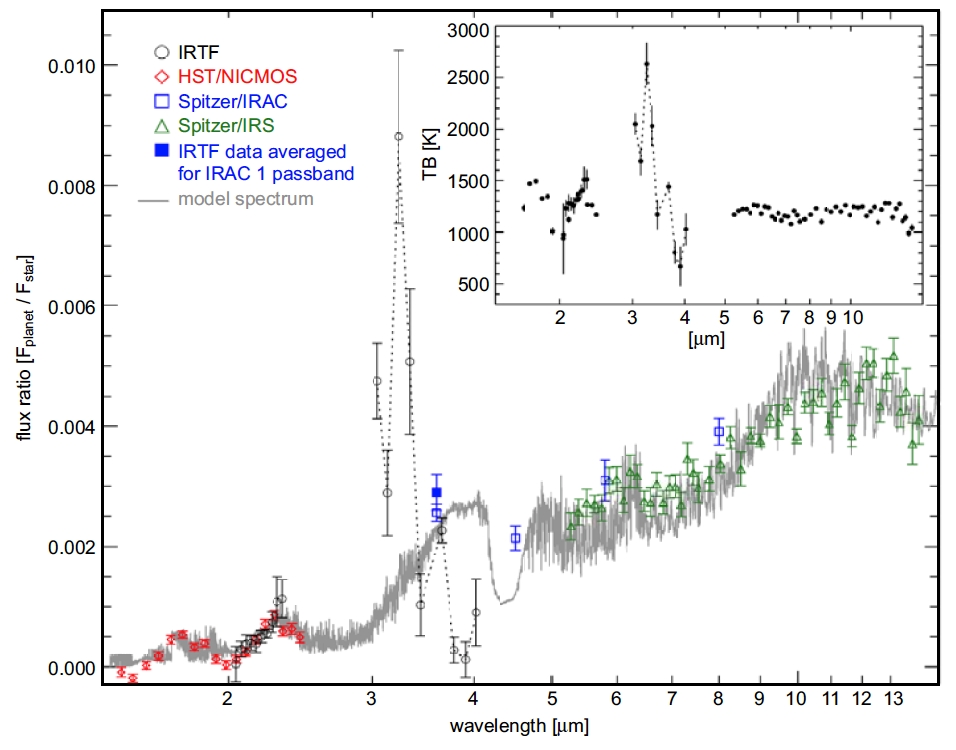}
\caption{\label{fig5} (reproduced from fig. 2 in \cite{Swain2010})
  Emission spectrum of HD189733b observed with the 3,0-m IRTF (open
  circles), compared to previous HST observations (red), SST
  observations (blue and green), and a radiative transfer model (grey
  solid line) assuming LTE and adjusted to the HST and SST data.}
\end{figure}

\section{Summary}

The detection and characterization of exoplanet amospheres is a brand
new field which has started to bloom just in the past 3--4 years,
thanks first to the advent of the SST, and most recently to the quick
start off of successful detections with telescopes on the
ground. Basically all atmospheres detected thus far are of hot
Jupiters, but these detections are just the tip of the iceberg for
what is yet to come. In the next few years, with more advanced
instruments coming online, and the rapid improvement of analysis
techniques, the characterization of exoplanet atmospheres will soon
start to advance towards the lower mass planets regime, with the final
goal in mind being the detection and characterization of Earth-like
atmospheres.


\begin{thebibliography}{}
\small
%
%
%
\bibitem{Wolszczan1992}{Wolszczan, A., \& Frail, D.A. 1992, Nature, 355, 145}
\bibitem{Wolszczan1994}{Wolszczan, A. 1994, Science, 264, 538}
\bibitem{Mayor1995}{Mayor, M., \& Queloz, D. 1995, Nature, 378, 355}
\bibitem{Charbonneau2000}{Charbonneau, D., Brown, T.M., Latham, D.W.,
  \& {Mayor}, M. 2000, ApJL, 529, L45}
\bibitem{Henry2000}{Henry, G.W., Marcy, G.W., Butler, R.P., \& Vogt,
  S.S. 2000, ApJL, 529, L41}
\bibitem{Bond2004}{Bond, I. A. et al. 2004, APJL, 606, L155}
\bibitem{Kalas2008}{Kalas, P., et al. 2008, Science, 322, 1345}
\bibitem{Marois2008}{Marois, C., et al. 2008, Science, 322, 1348}
\bibitem{Spiegel2010}{Spiegel, D.S \& Burrows, A. 2010, ApJ, 722, 871}
\bibitem{Janson2010}{Janson, M., Bergfors, C., Goto, M., Brandner, W.,
  \& Lafreni\`ere, D. 2010, ApJL, 710, L35}
\bibitem{Lopez2007}{L\'opez-Morales, M., \& Seager, S. 2007, ApJL, 667, L191}
\bibitem{Charbonneau2002}{Charbonneau, D., Brown, T.M., Noyes, R.W., \& {Gilliland}, R.L. 2002, ApJ, 568, 377}
\bibitem{Tinetti2007}{Tinetti, G., et al. 2007, Nature, 448, 169}
\bibitem{Swain2008}{Swain, M.R., Vasisht, G., \& Tinetti, G. 2008,
  Nature, 452, 329}
\bibitem{Deming2005}{Deming, D., Seager, S., Richardson, L.J. \& Harrington, J. 2005,
  Nature, 434, 740}
\bibitem{Charbonneau2005}{Charbonneau, D.,  et al. 2005,
  ApJ, 626, 523}
\bibitem{Harrington2006}{Harrington, J., et al. 2006, Science, 314, 623}
\bibitem{Knutson2007}{Knutson, H.A., et al. 2007, Nature, 447, 183}
\bibitem{Fortney2008}{Fortney, J.J., Lodders, K., Marley, M.S., \& Freedman, R.S. 2008, ApJ, 678, 1419}
\bibitem{Zahnle2009}{Zahnle, K., Marley, M.S., Freedman, R.S.,
  Lodders, K., \& Fortney, J.J. 2009, ApJL, 701, L20}
\bibitem{Snellen2009}{Snellen, I.A.G., de Mooij, E.J.W., \& Albrecht,
  S. 2009, Nature, 459, 543}
\bibitem{Snellen2010}{Snellen, I.A.G., de Mooij, E.J.W., \& Burrows,
  A. 2010, A\&A, 513, A76 Nature, 452, 329}
\bibitem{Borucki2009}{Borucki, W. J. et al. 2009, Science, 325, 709}
\bibitem{Grillmair2007}{Grillmair, C.J., et al. 2007, ApJL, 658, L115}
\bibitem{Pont2008}{Pont, F., et al. 2008, MNRAS, 385, 109}
\bibitem{Grillmair2008}{Grillmair, C.J., et al. 2008, Nature, 456, 767}
\bibitem{Sing2009b}{Sing, D.K. et al. 2009, A\&A, 505, 891}
\bibitem{Redfield2008}{Redfield, S., Endl, M., Cochran, W.D., \&
      Koesterke, L. 2008, ApJL, 673, L87}
\bibitem{Snellen2008}{Snellen, I.A.G., Albrecht, S., de Mooij, E.J.W.,
  \& Le Poole, R.S. 2008, A\&A, 487, 357}
\bibitem{Snellen2010b}{Snellen, I.A.G., de Kok, R.J., de Mooij, E.J.W, Albrecht, S. 2010, Nature, 465, 1049}
\bibitem{Sing2010}{Sing, D.K. et al. 2010, ArXiv e-prints:1008.4795}
\bibitem{Colon2010}{Colon, K. D. etal. 2010, ArXiv e-prints:1008.4800}
\bibitem{Seager2000}{Seager, S., \& Sasselov, D.D. 2000, ApJ, 537, 916}
\bibitem{Sing2009a}{Sing, D.K., \& L\'opez-Morales, M. 2009, A\&A, 493, L31}
\bibitem{deMooij2009}{de Mooij, E.J.W., \& Snellen, I.A.G. 2009, A\&A, 493, L35}
\bibitem{Gillon2009}{Gillon, M., et al. 2009, A\&A, 506, 359}
\bibitem{Rogers2009}{Rogers, J.C., Apai, D., L\'opez-Morales, M.,
  Sing, D.K., \& Burrows, A. 2009, ApJ, 707, 1707}
\bibitem{Lopez2010}{L\'opez-Morales, M., et al. 2010, ApJ, 716, L36}
\bibitem{Hebb2009}{Hebb, L., et al. 2009, ApJ, 693, 1920}
\bibitem{Campo2010} {Campo, C. J. et al. 2010, ArXiv e-prints:1003.2763}
\bibitem{Husnoo2010}{Husnoo, N., et al. 2010, ArXiv e-prints:1004.1809}
\bibitem{Croll2010a}{Croll, B., et~al. 2010a, ArXiv e-prints:1009.0071}
\bibitem{Anderson2010}{Anderson, D. R. et al. 2010, A\&A, 513, L3}
\bibitem{Gibson2010}{Gibson, N. P. et al. 2010, MNRAS, 404, L114}
\bibitem{Croll2010b}{Croll, B., Albert, L., Lafreniere, D.,
  Jayawardhana, R., \& Fortney, J.J. 2010b, ApJ, 717, 1084}
\bibitem{Croll2010c}{Croll, B., Jayawardhana, R., Fortney, J.J.,
  Lafreni\`ere, D., \& Albert, L. 2010c, ApJ, 718, 920}
\bibitem{ODonovan2010}{O'Donovan, F.T. et al. 2010, ApJ, 710, 1551}
\bibitem{Mandell2010}{Mandell, A.M., et al. 2010, ArXiv e-prints:1011.5507}
\bibitem{Swain2010}{Swain, M.R., et al. 2010, Nature, 463, 637}
\bibitem{Charbonneau2008}{Charbonneau, D. et al. 2008, ApJ, 686, 1341}

\end{thebibliography}

%
%
\small  
%
%

%


\end{document}